\documentclass[12pt]{iopart}

\newcommand{\sNN}{$\sqrt{s_{NN}} = 2.76$ TeV }
\usepackage{graphicx}
\usepackage{epstopdf}
\begin{document}

\title{Untriggered di-hadron correlations in Pb-Pb collisions at $\sqrt{s_{NN}} =$ 2.76 TeV from ALICE  }

\author{Anthony R. Timmins for the ALICE collaboration}

\address{Department of Physics, University of Houston, 617 Science and Research, Building 1, Houston, TX 77204, USA}
\ead{anthony.timmins@cern.ch}
\begin{abstract}

We present measurements of untriggered di-hadron correlations as a function of centrality in Pb-Pb \sNN collisions, for charged hadrons with $p_{T} > 0.15$ GeV$/c$. These measurements provide a map of the bulk correlation structures in heavy-ion collisions. Contributions to these structures may come from jets, initial density fluctuations, elliptic flow, resonances, and/or momentum conservation. We decompose the measured correlation functions via a multi-parameter fit in order to extract the nearside Gaussian, the longer range $\Delta \eta$ correlation often referred to as the soft ridge. The effect of including higher harmonics ($v_{3}$ and $v_{4}$) in this procedure will be discussed. We investigate how the nearside Gaussian scales with the number of binary collisions. Finally, we show the charge dependence of the nearside Gaussian.
\end{abstract}



Measurements of untriggered di-hadron correlations aim to explore the bulk correlation structures in heavy-ion collisions. All charged hadrons with $p_{T} > 0.15$ GeV/c are used to form a correlation function, which for this analysis is defined as follows \cite{STAR1}:
\begin{equation}
\label{equ:2DCorr}
\frac{\Delta \rho}{\sqrt{ \rho_{ref}}} (\Delta \eta,\Delta \phi)  =  \frac{\rho_{sib}-\rho_{ref}}{\sqrt{ \rho_{ref}}} = \frac{d^2N_{ch}}{d\eta d\phi}(\frac{\rho_{sib}}{\rho_{ref}}-1)
\end{equation}
The symbol $\rho_{sib}$ refers to the number of correlated and uncorrelated pairs (or pair density), where the hadrons used to form the pair come from the same event.  $\rho_{ref}$ refers to the number of uncorrelated pairs, since the hadrons used to form the pair come from different events. $\Delta \rho $ is therefore the number of correlated pairs. The denominator is the square root of the number of background pairs, so that represents the number of charged hadrons. i.e $d^{2}N_{ch}/d\eta d \phi$. The correlation function therefore measures the \emph{number of correlated pairs per particle}.

\begin{figure*}[h]
\begin{center}
\includegraphics[width = 1\textwidth]{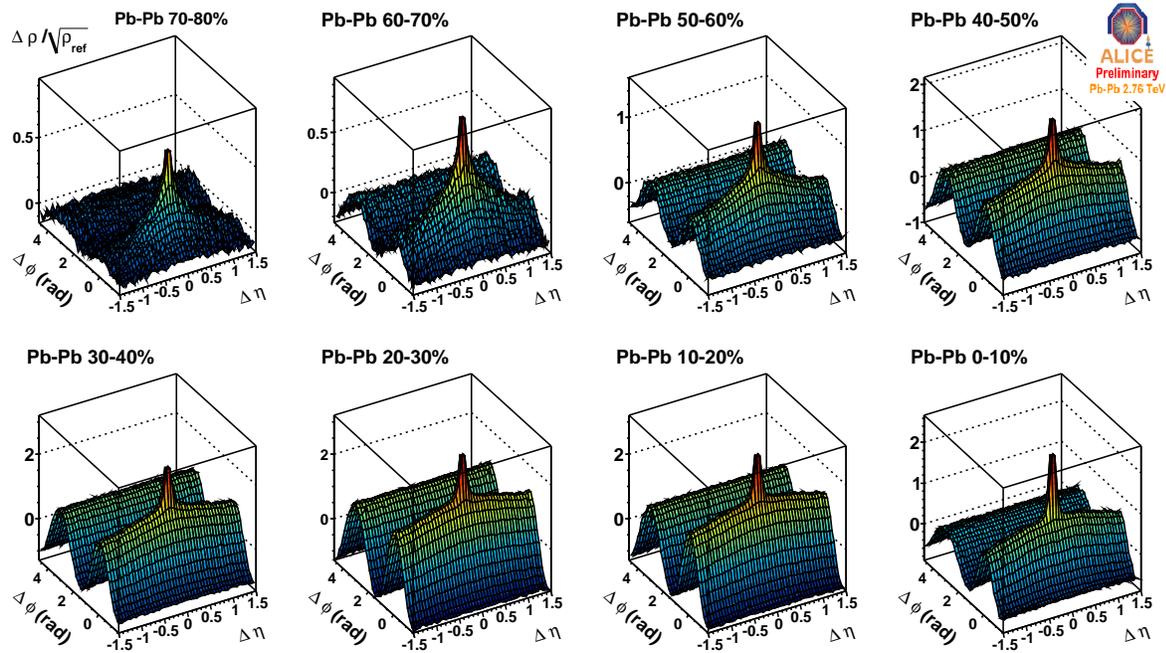}
\end{center}
\caption{Untriggered di-hadron correlations as a function of centrality in Pb-Pb \sNN collisions.}
\label{Fig1}
\end{figure*}

Figure \ref{Fig1} shows the extracted correlation function as a function of centrality. The amplitude of the longer range $\Delta \eta $ correlations grows from peripheral to mid-central collisions, then falls slightly for central collisions. On the nearside, the most prominent structure is a short range nearside spike, which sits on top of a longer range nearside Gaussian. The short range spike was found to be from HBT and $\gamma \rightarrow e^{+}+e^{-}$ conversions. Elliptic flow is also prominent, and is represented by the $\cos(2\Delta \phi)$ structure, which is independent of $\Delta \eta$. 

We attempt to characterise the longer range nearside Gaussian via a fit decomposition. This structure is of particular interest, since there are various interpretations with regard to its origin \cite{sr1, sr1i, sr2, sr3, sr4}, some of which give quantitative predictions at LHC energies. It is sometimes referred to as the ``soft ridge''. The fit decomposition is as follows:
\begin{eqnarray*}
\label{equ:Decomp2} 
\frac{\Delta \rho}{\sqrt{ \rho_{ref}}} (\Delta \eta,\Delta \phi) &=& Constant +  \textrm{Gauss}(\Delta \phi, \Delta \eta) + A\cos(\Delta \phi)+B\cos(2\Delta \phi) \\ \\ 
&& + C\cos(3\Delta \phi)+D\cos(4\Delta \phi)
\end{eqnarray*}
We extract the 2D Gaussian terms ($\textrm{Gauss}(\Delta \phi, \Delta \eta))$ under two scenarios: the first fixes $C=D=0$, the second allows $C$ and $D$ to free parameters, which in turn enables higher order harmonics to be included in the prescription. The $B\cos(2\Delta \phi)$ term originates from elliptic flow, while $A\cos(\Delta \phi)$ term may have a variety of origins e.g. momentum conservation, directed flow, and/or awayside jets. Bins in the region which covers a circle centered at 0,0 with a of radius 0.5 are given zero weight in the fit, to prevent the spike influencing the fit. The $\chi^2/dof$ for both scenarios varies between $1 \rightarrow 1.5$ and no remaining structures were found to exist in the 2D residuals.
\begin{figure*}[h]
\begin{center}
\includegraphics[width = 0.9\textwidth]{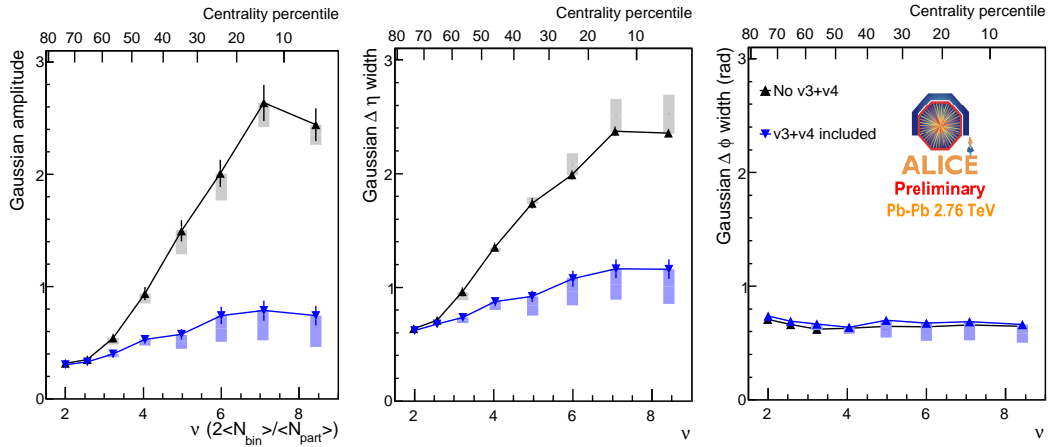}
\end{center}
\caption{Extracted Gaussian parameters as a function of centrality in Pb-Pb \sNN collisions. ``No v3+v4'' refers to fits without higher harmonics, ``v3+v4 included'' refers to fits with higher harmonics. Boxed errors indicate systematic uncertainties in the $\Delta \eta$ width of $\rho_{ref}$, line errors indicate systematic uncertainties in the normalisation of correlation function and extracted parameters in the fitting routine. Connecting lines guide the eye.}
\label{Fig2}
\end{figure*}
Figure \ref{Fig2} shows the extracted nearside Gaussian parameters. Previous analyses at RHIC energies negated the higher harmonic terms \cite{STAR1, Daugherity}, and we find the nearside Gaussian without higher harmonics behaves in a similar way: a substantial growth is seen for both the amplitude and the $\Delta \eta$ width. When higher harmonics are included, the behaviour of the Gaussian is more subdued with relatively modest increases in the amplitude and $\Delta \eta$ width. The reduction in the Gaussian terms is accompanied by increases in $B, C, D$ which are not shown. It is possible to convert the amplitudes $A,B, C, D$ to $v_n$ coefficients \cite{Kettler}, and these were found to be consistent with another analysis at low $p_{T}$ \cite{Adare}, for the case where $C$ and $D$ are free parameters.
\begin{figure*}[h]
\begin{center}
\includegraphics[width = 0.35\textwidth]{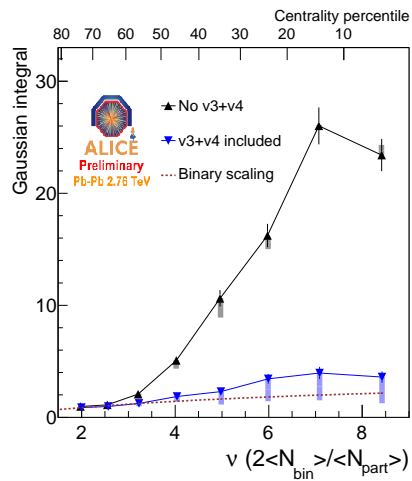}
\end{center}
\caption{Extracted Gaussian integral as a function of centrality in Pb-Pb \sNN collisions. See figure \ref{Fig2} caption for description of labels and errors. Binary scaling assumes the number of correlated pairs in the Gaussian scales with the number of binary collisions, with Pb-Pb 70-80\% as the reference. Connecting lines guide the eye.}
\label{Fig3}
\end{figure*}
Figure \ref{Fig3} shows the integral of the nearside Gaussian. The dashed lines assumes the number of correlated pairs in the Gaussian scales with the number of binary collisions from peripheral to more central collisions. It appears that this hypothesis is favoured more by the Gaussian extracted with higher harmonics in the fit. Finally, figure \ref{Fig4} shows the charge dependence of the nearside Gaussian. 
\begin{figure*}[h]
\begin{center}
\includegraphics[width = 0.85\textwidth]{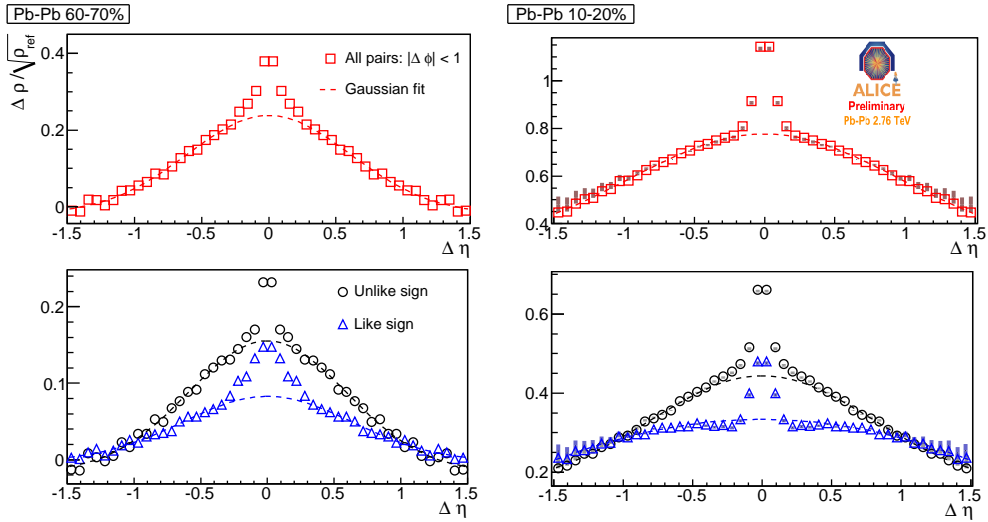}
\end{center}
\caption{$\Delta \eta$ projection of the near side for different pair charge signs in Pb-Pb \sNN collisions. Boxed errors are described earlier, line errors indicate systematic uncertainties in normalisation of correlation function. Gaussian fits are done in the $|\Delta \eta| > 0.5$ region. }
\label{Fig4}
\end{figure*}
It is clear that unlike sign pairs contribute more to the near structure within ALICE's acceptance. Like sign pairs have a broader structure. If the nearside Gaussian were driven only by a flow mechanism which acts on charged hadrons independent of charge sign, one would expect no charge dependence, which is in contrast to our data.

In summary, we have measured untriggered di-hadron correlations in Pb+Pb \sNN collisions. A pronounced change in the correlation structure from peripheral to central collisions is observed. We have quantified the nearside Gaussian with two methods: including and omitting higher harmonics which has a significant effect on the extracted Gaussian parameters.
The extracted Gaussian integral with higher harmonics scales with the number of binary collisions, within the quoted uncertainties. Finally, we observe a charge sign dependence for the nearside Gaussian: unlike sign correlations are narrower and stronger within ALICE's acceptance.

\end{document}